\documentclass[conference]{IEEEtran}
\IEEEoverridecommandlockouts
\usepackage{cite}
\usepackage{amsmath,amssymb,amsfonts}
\usepackage{algorithmic}
\usepackage{graphicx}
\usepackage{textcomp}
\usepackage{xcolor}
\def\BibTeX{{\rm B\kern-.05em{\sc i\kern-.025em b}\kern-.08em
    T\kern-.1667em\lower.7ex\hbox{E}\kern-.125emX}}

\usepackage{booktabs}
\usepackage{comment}
\usepackage{subcaption}
    
\begin{document}

\title{Applying an Evolutionary Algorithm to Minimize Teleportation Costs in Distributed Quantum Computing
\thanks{© 2024 IEEE.  Personal use of this material is permitted.  Permission from IEEE must be obtained for all other uses, in any current or future media, including reprinting/republishing this material for advertising or promotional purposes, creating new collective works, for resale or redistribution to servers or lists, or reuse of any copyrighted component of this work in other works.

Sponsored in part by the Bavarian Ministry of Economic Affairs, Regional Development and Energy as part of the 6GQT project.}
}

\author{\IEEEauthorblockN{Leo Sünkel}
\IEEEauthorblockA{
\textit{LMU Munich}\\
Munich, Germany \\
leo.suenkel@ifi.lmu.de}
\and
\IEEEauthorblockN{Manik Dawar}
\IEEEauthorblockA{
\textit{LMU Munich}\\
Munich, Germany \\
m.dawar@physik.uni-muenchen.de}
\and
\IEEEauthorblockN{Thomas Gabor}
\IEEEauthorblockA{
\textit{LMU Munich}\\
Munich, Germany \\
thomas.gabor@ifi.lmu.de}
}

\maketitle

\begin{abstract}
By connecting multiple quantum computers (QCs) through classical and quantum channels, a quantum communication network can be formed. This gives rise to new applications such as blind quantum computing, distributed quantum computing, and quantum key distribution. In distributed quantum computing, QCs collectively perform a quantum computation. As each device only executes a sub-circuit with fewer qubits than required by the complete circuit, a number of small QCs can be used in combination to execute a large quantum circuit that a single QC could not solve on its own. However, communication between QCs may still occur. Depending on the connectivity of the circuit, qubits must be teleported to different QCs in the network, adding overhead to the actual computation; thus, it is crucial to minimize the number of teleportations. In this paper, we propose an evolutionary algorithm for this problem. More specifically, the algorithm assigns qubits to QCs in the network for each time step of the circuit such that the overall teleportation cost is minimized. Moreover, network-specific constraints such as the capacity of each QC in the network can be taken into account. We run experiments on random as well as benchmarking circuits and give an outline on how this method can be adjusted to be incorporated into more realistic network settings as well as in compilers for distributed quantum computing. Our results show that an evolutionary  algorithm is well suited for this problem when compared to the graph partitioning approach as it delivers better results while simultaneously allows the easy integration and consideration of various problem-specific constraints.
\end{abstract}


\section{Introduction}
Multiple quantum computers (QCs) may be connected with each other through classical and quantum channels, forming quantum communication networks (QCNs); from these, the so-called quantum internet may arise \cite{kimble2008quantum,kozlowski2019towards,awschalom2021development} with the potential of ushering in a new communication era. This allows for new types of applications in quantum computing, including distributed quantum computing (DQC)~\cite{cirac1999distributed}, blind quantum computing (BQC) \cite{fitzsimons2017private}, and quantum key distribution (QKD) \cite{bennett1984quantum}. While BQC and QKD protocols provide advantages in both security and privacy, DQC provides the possibility of running large quantum circuits that may require too many qubits for a single QC. In the current NISQ-era \cite{preskill2018quantum}, QCs consist of relatively few and noisy qubits, and these restrictions may remain for the foreseeable future. Moreover, it remains to be seen how far a single QC can be scaled. In this regard, DQC may offer a viable path to scaling QCs to include more qubits \cite{caleffi2022distributed}. 
However, QCNs do not only offer new opportunities, they also come with novel problems and challenges. For instance, entanglement must be generated, distilled and distributed to enable long distance communication. Another example is in the domain of DQC, where new compilers are required \cite{ferrari2021compiler}. Communication between quantum nodes in the network cause overhead, as communication in a quantum network is performed through the use of quantum teleportation, which entails entanglement, a valuable resource in quantum communication.
In DQC, circuits are divided into sub-circuits that then can each be executed separately on smaller QCs or quantum processing units (QPUs). However, qubits must be teleported to different devices throughout the network whenever a multi-qubit gate whose qubits are stored on different devices must be executed. As this communication causes some overhead and consumes costly resources, the number of teleportations should be kept at a minimum, i.e., should be minimized as much as possible. This is the problem which we focus on in this paper, that is, minimizing the teleportation cost between quantum nodes for a given quantum circuit. For this we propose an evolutionary algorithm (EA) that takes as input a quantum circuit and assigns each qubit to a QPU, which can then execute a sub-circuit individually. Additionally, the algorithm allows easy integration of various constraints such as number of QPUs (i.e. nodes in a quantum network) as well as the capacity of each QPU. We run experiments on random as well as known benchmarking circuits and compare the results of our EA to a graph partitioning approach.

This paper is structured as follows. In Section \ref{sec:background} we briefly recap the fundamentals of quantum computing, quantum communication networks, distributed quantum computing and evolutionary algorithms. In Section \ref{sec:related_work} we discuss related work to our approach while we discuss the details of our proposed algorithm in Section \ref{sec:algorithm}. The results of our experiments are discussed in Section \ref{sec:experiments}. We conclude the paper and give an outlook in Section \ref{sec:conclusion}.

\section{Background}\label{sec:background}
We discuss quantum computing, quantum communication networks (QCNs), distributed quantum computing (DQC) and evolutionary algorithms (EAs) in this section.
\paragraph*{\textbf{Quantum Computing}}
The most fundamental building block in quantum computing is the \textit{qubit}, whose counterpart is the \textit{bit} in classical information theory. A qubit can be defined as $|\psi\rangle = \alpha|0\rangle + \beta|1\rangle$ where $\alpha$ and $\beta$ are complex numbers and $|\alpha|^2$ and $|\beta|^2$ sum up to 1. The basis vectors $|0\rangle$ and $|1\rangle$  represent states 0 and 1 respectively. While both bits and qubits are two-state systems, a crucial difference is that a qubit can be in a \textit{superposition}, i.e., a linear combination of states. More specifically, in the the definition above, $|\alpha|^2$ and $|\beta|^2$ represent the probabilities of the qubit collapsing to a particular state upon \textit{measurement}, i.e., the act of measurement causes the quantum state to collapse, at which point the qubit takes on a classical value (0 or 1). Until measurement, the state of the qubit is unknown and it may appear as being in several states simultaneously \cite{nielsen2002quantum}.
Multiple qubits can become \textit{entangled}, i.e., their measurement outcomes correlate, even if the qubits are physically far apart.
A well-known example of an entangled state is the \textit{bell state} $|\psi\rangle = \frac{|00\rangle + |11\rangle}{\sqrt{2}}$. This is a maximally entangled state and plays a vital role in QCNs.
While classical information can easily be copied, quantum information cannot be copied at all as this is prohibited by the \textit{no cloning theorem}. However, quantum information, i.e., quantum states, can be transferred through the use of a \textit{teleportation} protocol \cite{bennett1993teleporting}. Note that only quantum information is transferred (i.e. no physical teleportation is involved) and the original quantum state is destroyed in the process, complying with the no cloning theorem. Teleportation is a crucial building block in QCNs and we will elaborate on this in the next section.
\paragraph*{\textbf{Quantum Communication Networks}}
A QCN can be defined as a network consisting of quantum nodes that are connected via classical and quantum channels \cite{kimble2008quantum,kozlowski2019towards,van2013designing}. Any quantum communication between a pair of nodes requires shared entanglement pairs among them. This makes entanglement the fundamental resource of a quantum network. There are numerous protocols that can enable communication. The most important one is teleportation, which enables the transfer of the quantum state of a qubit at one node to a qubit stored at another node. Quantum teleportation is a process by which the state of a qubit can be transmitted from one location to another, with the help of an entangled qubit pair shared between the sender (Alice) and the receiver (Bob). The process of quantum teleportation involves three main steps:

\begin{enumerate}
    \item Alice and Bob share an entangled pair of qubits.
    \item Alice performs a Bell measurement on her qubit and the qubit she wants to teleport, obtaining two classical bits of information.
    \item Bob uses these two classical bits to perform one of four operations -- a Pauli $X$, a Pauli $Z$, or both Pauli $X$ and $Z$ -- on his half of the entangled pair, obtaining the original quantum state.
\end{enumerate}

The important thing to note is that each teleportation consumes an entanglement pair shared between Alice and Bob. 

As quantum channels are noisy and nodes may be separated far apart, so called \textit{quantum repeaters} \cite{briegel1998quantum,van2013designing} are stationed throughout the network. The role of a quantum repeater is to enable the establishment of entanglement of nodes (via the entanglement of qubits) that are spatially separated through a process known as \textit{entanglement swapping} \cite{van2013designing}. This protocol requires two Bell pairs and three parties, who we will call Alice, Bob, and Charlie. Alice and Charlie each possess a qubit of the same Bell pair and Charlie and Bob each also share a qubit of another Bell pair. Note that at this point Alice and Bob do not share a Bell pair. However, through the application of the protocol, entanglement between Alice and Bob can be established. The entanglement fidelity can further be increased with a quantum repeater by employing \textit{entanglement purification} or \textit{entanglement distillation}\cite{zukowski1993event,briegel1998quantum,van2013designing}.

\paragraph*{\textbf{Distributed Quantum Computing}}
QCNs allow for several new quantum computing applications and paradigms, one of which is DQC \cite{caleffi2022distributed}. In DQC, quantum circuits are divided into a number of sub-circuits, which can then be distributed throughout the network and executed simultaneously by the different nodes available in the network. This allows, for example, the execution of large quantum circuits that require too many qubits for a single quantum computer. 
Circuits must be partitioned, i.e., qubits are assigned to a QPU. It is often inevitable that multi-qubit gates (e.g. CNOT) are partitioned in such a way that the control and target qubits are located in different nodes. In this situation, one of the qubits must either be teleported to the other node (TeleData), or a \textit{remote} gate (TeleGate) must be applied across the two nodes. In either case, one entanglement pair is consumed. Since entanglement is the fundamental resource of a quantum network, it is in our best interests to utilize it efficiently. Each teleportation (of a qubit or gate) incurs a cost -- an entanglement pair -- which needs to be minimized.
\paragraph*{\textbf{Evolutionary Algorithms}}
The field of evolutionary computing \cite{eiben2002evolutionary} describes meta-heuristic optimization algorithms such as EAs or genetic algorithms (GAs) that are loosely inspired by biological evolution and can be applied to optimization problems from various domains. We use the term evolutionary algorithm (EA) in this paper. 
Solutions to a particular problem are encoded by \textit{individuals} and a collection of individuals form a \textit{population}. Each individual has a \textit{fitness} associated with its solution quality, i.e., a numeric value that describes how ``good'' a particular solution is. Thus the aim of the algorithm is to maximize this fitness value. The fitness itself is determined by a problem specific \textit{fitness function}, i.e., a function that takes a solution to a problem as input and returns a numeric value corresponding to the goodness of said solution.
The evolutionary process is iterative, that is, the EA runs for a number of \textit{generations} or until a termination criterion is met. In each generation, new solutions (``children'') are created by a process known as \textit{crossover}, which combines parts of the solutions from two existing individuals (``parents''). 
For example, in the single point crossover approach, a random cut-off point is selected. A child is then created by inheriting parts of the solution left to the cut-off point from one parent while the rest is acquired of the solution right to the cut-off point from the other parent.
Another example is the uniform crossover approach in which each variable is inherited from either parent with an equal probability.
After crossover, random mutations, i.e., small changes to a solution, may be applied. An example mutation operation is the bit-flip  operation in which a random bit is selected from the solution and flipped (i.e., 0 is changed to 1 and vice versa).
At the end of each generation, \textit{survivor selection} takes place, which is the process of determining what solutions are allowed to be part of the next generation and what solutions are discarded. Selection is largely based on an individuals fitness value, however, different survivor selection strategies have been introduced in the literature.

To summarize the entire approach: an EA initially begins with a population initialized with random solutions (i.e. individuals). For a number of generations, new solutions are created by randomly combining parts of different solutions in a process known as crossover. These solutions can further be modified by random mutations. At the end of each generation, a new population is created by selecting individuals deemed fit and the process repeats in the next generation. Over time, ``bad solutions'' are discarded and ``good'' solutions thrive. However, EAs are based on randomness and thus convergence is not guaranteed, i.e., the best solution is not necessarily found in each run of the algorithm.
\section{Related Work}\label{sec:related_work}
Many approaches for quantum circuit compilation for DQC have been proposed in recent years. Quantum circuit compilation can be divided into three sub-problems: partitioning, remote-gate scheduling, and local routing.
In \cite{andres2019automated}, the authors focus on partitioning. They propose to represent a quantum circuit as a hypergraph where the nodes are the qubits and CZ-gates and the hyperedges are groups of CZ gates. They then use a state of the art hypergraph partitioning solver \cite{kahypar} to partition the hypergraph and, hence, the quantum circuit. This is done to minimize the number of inter-QPU gates.
\cite{sundaram2023} proposes two polynomial-time heuristics for partitioning quantum circuits: Local-Best and Zero-Stitching. 
Local-Best starts with a good initial assignment of the qubits to the different partitions. Then, as the circuit is read in time order, for any encountered non-local two-qubit gate, a best assignment of its operands is determined by assessing the other gates they participate in in the near future. 
Zero-stitching divides the original circuit time-wise, into a number of zero-cost sub-circuits. Then these circuits are stitched wherever needed for the successful execution of the circuit.
Assuming we are only considering one- and two-qubit gates, then once the partitions are determined, however, there is still the problem of deciding, for each remote gate, whether to teleport the target or the control qubit. \cite{moghadam2016} tries to answer this question: The paper proposes an exponential time algorithm to optimally decide the teleportations of the qubits once the partitions are already assigned. 
\cite{houshmand2020evolutionary} proposes a genetic algorithm to speed up the brute-force approach used in \cite{moghadam2016}.
\cite{ferrari2023modular} proposes approaches that deal with all the three above mentioned problems of quantum circuit compilation: partitioning, remote-gate scheduling and local routing. 
First a qubit assignment is determined using METIS's multilevel $k$-way partitioning, where it is assumed that all QPUs are of roughly the same size. Since this is likely suboptimal, the authors iteratively improve upon the output from the $k$-way partitioning by moving individual qubits around and accepting only those modifications that reduce the number of teleportations. 
Then remote gate scheduling is performed by scanning through all the non-local two-qubit gates and then, for each such gate, determining the optimal TeleData or TeleGate operation that would require a minimal number of teleportations to execute. 
Then finally, for local routing, first the non-local two-qubit gates are identified. For each such gate, for its operands, the shortest sequence of SWAP gates is determined for them to reach their nearest communication qubits, so they can then be teleported. 
\cite{dadkhah2021new} proposes a genetic algorithm for partitioning. But unlike our approach, the authors do not take into account the time-dimension of the circuit, i.e., the fact that a circuit's connectivity can change significantly from layer to layer.

\section{An Evolutionary Algorithm to Minimize Teleportations}\label{sec:algorithm}
Recall that all qubits of a multi-qubit gate must be located in the same node (QPU). This can, for instance, be achieved by teleportation, i.e., the qubits' state is transferred to the respective node. However, as this implies some overhead, the goal to minimize such overhead, and thus the teleportation cost, naturally arises. In this section, we introduce our proposed algorithm to minimize the number of teleportations of a distributed quantum circuit.

\paragraph*{\textbf{Solution Representation}}
We represent a single solution as an $n \times m$ matrix, where $n$ is the number of qubits and $m$ the number of time steps in a given circuit. Each cell in this matrix contains an integer value that specifies in what partition the particular qubit is at the respective point in time. Thus, this representation allows for an arbitrary number of QPUs and takes the time domain as well as the dynamics of the circuit into account. 

Often a qubit is immediately teleported back to its original partition once the multi-qubit gate has been executed. However, in the representation here this is not necessarily the case as it may remain in a particular partition until it is required elsewhere. 

\begin{figure}[tb]
    \centering
    \includegraphics[scale=0.5]{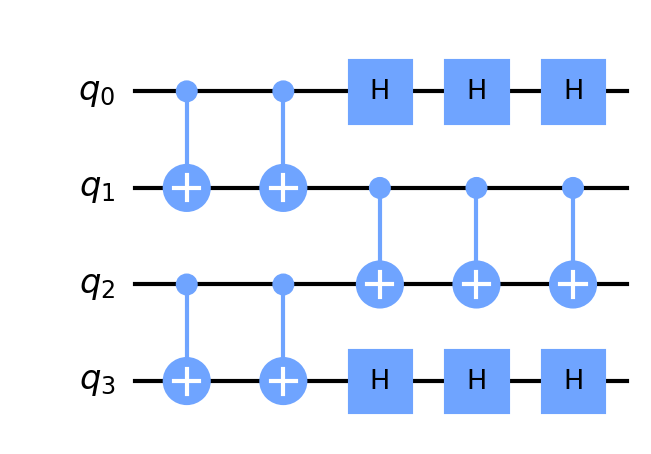}
    \caption{Example circuit. Note that this circuit is for illustration purposes only, it is by no means meant as a realistic circuit. Circuit created with Qiskit \cite{Qiskit}}
    \label{fig:example_circuit}
\end{figure}

An example circuit is shown in Fig. \ref{fig:example_circuit} and a corresponding solution representation can bee seen in Table \ref{tab:solution_representation_with_time}. In this solution, qubits $q_0$ and $q_1$ are assigned to QPU 1 in time steps 0 and 1 while $q_2$ and $q_3$ are assigned to QPU 0. After this, two teleportations occur and qubits $q_0$  and $q_3$ are assigned to QPU 0 while qubits $q_1$ and $q_2$ are assigned to QPU 1.

\begin{table}[tb]
\caption{Solution representation}
\label{tab:solution_representation_with_time}
    \centering
    \begin{tabular}{|c|c|c|c|c|c|}
    \hline
    \multicolumn{6}{|c|}{\textbf{Time}} \\ \hline
        $t_i / q_i$ & $t_0$ & $t_1$ & $t_2$ & $t_3$ & $t_4$ \\ \hline
        $q_0$ & 1 & 1 & 0 & 0 & 0 \\ \hline
        $q_1$ & 1 & 1 & 1 & 1 & 1 \\ \hline
        $q_2$ & 0 & 0 & 1 & 1 & 1 \\ \hline
        $q_3$ & 0 & 0 & 0 & 0 & 0 \\ \hline 
    \end{tabular}
\end{table}

\begin{table}[tb]
\caption{Solution representation}
\label{tab:solution_representation_with_time_extended}
    \centering
    \begin{tabular}{|c|c|c|c|c|c|c|c|}
    \hline
    \multicolumn{8}{|c|}{\textbf{Time}} \\ \hline
        $t_i / q_i$ & $t_0$ & $t_1$ & $t_2$ & $t_3$ & $t_4$ & $t_5$ & $t_6$ \\ \hline
        $q_0$ & 1 & 1 & 0 & 0 & 0 & 1 & 1\\ \hline
        $q_1$ & 1 & 1 & 1 & 1 & 1 & 1 & 1 \\ \hline
        $q_2$ & 0 & 0 & 1 & 1 & 1 & 0 & 0 \\ \hline
        $q_3$ & 0 & 0 & 0 & 0 & 0 & 0 & 0\\ \hline 
    \end{tabular}
\end{table}

\paragraph*{\textbf{Fitness Function}}
The overall goal is to minimize the number of teleportations between nodes. However, some constraints must be fulfilled. More specifically, a node cannot be assigned more qubits at any given time than allowed by its capacity and qubits of the same multi-qubit gate must be teleported to the same partition. In the fitness function that was applied in this work, the fitness was determined in the following way:

\begin{enumerate}
    \item Add a teleportation cost of 1 once the partition of a qubit changes from one time step to the next
    \item  Add a teleportation cost of 1 if the qubits of a multi-qubit gate are not in the same partition. 
    \item Add a penalty $\delta$ (100 in our experiments) if the capacity of a node is exceeded.
\end{enumerate}

These values, i.e., costs and penalty terms, can be adjusted and new constraints can easily be enforced by adding new penalty terms. This approach also allows, for instance, to consider QPUs with vastly different capacity. However, for this study we kept the fitness function simple; for a more realistic setting more advanced functions may be required and merit further investigation.  

Consider the solution depicted in Table \ref{tab:solution_representation_with_time_extended}. This solution corresponds to a fitness value of 4 as the partitions for a qubit change a total of 4 times and for each partition change teleportations are required. In this solution no constraints are violated.

\paragraph*{\textbf{Crossover and Mutation}}
We apply two crossover and three mutation functions in our proposed algorithm. Children are created by single-point or uniform crossover, though new children may also be created randomly. In single-point crossover, a random cut-off point is selected and all parts to the left of this point are taken from one parent while the parts to the right of this point are taken from another. In uniform crossover, each column is individually chosen from random parent. Once a new solution has been created, it is subjected to random mutations. In the first mutation method, a partition is randomly assigned. In the second mutation, a subset of entries of a row in the solution representation above is shuffled while in the third method a subset of entries of a column is shuffled. More elaborate crossover and mutation operations can easily be incorporated to further improve the performance and results.

\section{Experiments}\label{sec:experiments}
In this section, we present and discuss the experimental results. We focus on a setting using two QPUs, each with an equal capacity. We compare our approach to the min cut on the resulting partitions resulting from a graph partitioning algorithm.

\paragraph*{\textbf{Experimental Setup}}
In our first set of experiments, the circuits were created randomly using Qiskit \cite{Qiskit}. We created circuits with a depth of 10,  20, 30, 40 and 50; however, circuits were decomposed multiple times after creation, so the used circuits may contain vastly more gates, the actual depth of the circuits is presented alongside the results in Table \ref{tab:random_circuits_results_avg_per_depth} and Table \ref{tab:qft_results_avg_per_qubit}. We used circuits with 4 and 8 qubits. 
In our second set of experiments, we used circuits from MQT Bench \cite{quetschlich2023mqtbench}. That is, we used the QFT circuits with 4, 8, 16, 32, and 50 qubits. Note that the decomposed circuits are used in all experiments. The exact depth for each experiment is discussed below. We compare our EA to a graph partitioning (GP) algorithm. More specifically, we used the Kernighan Lin algorithm to obtain the partitions and applied the min cut function, both provided by NetworkX \cite{hagberg2008exploring}, to obtain the teleportation cost.

\begin{table}[tb]
\caption{Parameters used for the EA}
\label{tab:parameters}
    \centering
    \begin{tabular}{lc}
    \toprule
       Population size & 400 \\ \midrule
       Generations & 1000 \\ \midrule
       Crossover rate & 0.8 \\ \midrule
       Mutation rate & 0.8 \\ \midrule
       $\delta$ & 100 \\
    \bottomrule
    \end{tabular}
\end{table}

\paragraph*{\textbf{Random Circuits}}
For each circuit configuration (i.e. depth), a different circuit was randomly created, and the evolutionary algorithm was run 5 times, each time with a different seed. Thus, in total 5 circuits were used, and for each one the algorithm was executed using a different seed. A selection of the relevant parameters are listed in Table \ref{tab:parameters} and the average results of all circuits of a particular depth over all seeds is listed in Table \ref{tab:random_circuits_results_avg_per_depth}.

The EA performs better on average than GP for every circuit, and the difference in performance is clearly visible in the larger circuits. However, it is important to stress that in GP it is assumed that a qubit is teleported back to its original location after it has been teleported for a multi-qubit gate operation. It is equally important to stress that the performance of the EA is dependent on its parameters, that is, running the algorithm for more generations with a larger population may increase its performance even further. Adjusting crossover and mutation rate as well as implementing more problem-specific mutation and crossover operations could also lead to better results.

\begin{table}[tb]
\caption{Average results of random circuits over seeds.}
\label{tab:random_circuits_results_avg_per_depth}
    \centering
    \begin{tabular}{llll}
    \toprule
       \textbf{Qubits} & \textbf{Circuit depth} & \textbf{EA result} &  \textbf{GP result} \\ \midrule
       4 & 88  & 24.0 & 28.0 \\ \midrule
       4 & 146 & 36.0 & 46.0 \\ \midrule
       4 & 199 & 54.2 & 63.0 \\ \midrule
       4 & 241 & 62.8 & 74.0 \\ \midrule
       4 & 283 & 76.8 & 91.0 \\ \midrule
       8 & 99 & 21.4 & 36.2 \\ \midrule
       8 & 178 & 56.2 & 75.2 \\ \midrule
       8 & 301 & 93.4 & 132 \\ \midrule
       8 & 390 & 126.4 & 159 \\ \midrule
       8 & 511 & 161.8 & 209.4 \\ 
    \bottomrule
    \end{tabular}
\end{table}

\paragraph*{\textbf{Quantum Fourier Transformation Circuits}}
The circuits in these experiments were created with MQT Bench \cite{quetschlich2023mqtbench} with 4, 8, 16, 32 and 50 qubits. Note that the circuits were decomposed using Qiskit and the resulting depth is shown in Table \ref{tab:qft_results_avg_per_qubit} below. Results are also shown in the same table.
In the smallest circuit, the EA and GP approaches achieve the same results, however, for all remaining circuits, the EA outperforms the baseline algorithm.

\begin{table}[b]
\caption{Average results of the QFT circuits over seeds.}
\label{tab:qft_results_avg_per_qubit}
    \centering
    \begin{tabular}{llcc}
    \toprule
       \textbf{Qubits} & \textbf{Depth} & \textbf{EA result} &  \textbf{GP result} \\ \midrule
       4 & 26 & 8.0  & 8.0 \\ \midrule
       8 & 58 & 26.6 & 32.0 \\ \midrule
       16 & 122 & 119.4 & 128.0  \\ \midrule
       32 & 250 & 503.4 & 512.0 \\ \midrule
       50 & 394 & 1227.6 & 1253.0 \\ 
    \bottomrule
    \end{tabular}
\end{table}

\section{Conclusion}\label{sec:conclusion}
Connecting multiple quantum computers to form a QCN allows the implementation of the envisioned paradigm of DQC. However, the management of such a large networks provides new obstacles that need to be overcome as well as new problems to be solved. One such optimization problem is the minimization of teleportation cost between QPUs. We proposed an EA to solve this task while also considering network-specific constraints such as QPU capacity and number of available QPUs in the network. We ran experiments on randomly created circuits as well as QFT benchmarking circuits with varying size and compared our EA to an approach based on GP. The EA provides good initial results, but further investigation is required as both methods, EA and GP, rely on certain assumptions.
The solution encoding of the proposed algorithm takes the entire circuit into account, that is, each qubit is assigned to a specific QPU at each point in time, i.e., for each gate. This means that qubits are not necessarily teleported back to their original location, which may lead to better results. However, this encoding could also be improved by, for example, removing certain single qubit gates and thereby reducing the solution space, and further pre-processing steps may be envisioned.
An advantage of our proposed method is that multiple constraints can easily be integrated into the fitness functions. This allows specific details of a particular network to be taken into account: for instance, while in the experiments on random circuits we used two QPUs with a constant capacity, our approach allows for an arbitrary number of QPUs each with different capacities; this may be closer to a realistic setting. Furthermore, the proposed algorithm could be part of a larger compilation framework for DQC. Other network-specific constraints may be integrated; however, this requires further research.

\bibliographystyle{IEEEtran}
\bibliography{IEEEabrv,bibliography}

\end{document}